\documentclass[showpacs,preprintnumbers,amsmath,amssymb]{revtex4}
\usepackage{graphicx}
\usepackage{dcolumn}
\usepackage{bm}
\begin{document}
\title{Gauged Floreanini-Jackiw type chiral boson and its BRST
quantization}
\author{Anisur Rahaman} \email{1. anisur.rahman@saha.ac.in, 2.
manisurn@gmail.com} \affiliation{Hooghly Moghin College,
Chinsurah, Hooghly-712101, West Bengal, India}
\author{Safia Yasmin}
\affiliation{Indas Mahavidyalaya, Bankura - 722205, West Bengal,
India}

\date{\today}

\begin{abstract}
The gauged model of Siegel type chiral boson is considered. It has
been shown that the action of gauged model of Floreanini-Jackiw
(FJ) type chiral boson is contained in it in an interesting
manner. A BRST invariant action corresponding to the action of
gauged FJ type chiral boson has been formulated using Batalin,
Fradkin and Vilkovisky based improved Fujiwara, Igarishi and Kubo
(FIK) formalism. An alternative quantization of the gauge
symmetric action has been made with a Lorentz gauge  and an
attempt has been made to establish the equivalence between the
gauge symmetric version of the extended phase space and original
gauge non-invariant version of the usual phase space.
\end{abstract}

 \maketitle

\section{{\bf Introduction}}
The self-dual field in $(1+1)$ which is also known as chiral boson
is the basic ingredient of heterotic string theory \cite{ST1, ST2,
ST3, ST4}. This very chiral boson plays a crucial role in the
study of quantum hall
 effect too \cite{HAL1, HAL2}. Seigel initiated the study of chiral boson in his
 seminal work \cite{SIG}. Another description of chiral boson came from the
 work of Srivastva \cite{PPS}. In these two descriptions  \cite{SIG, PPS}, the lagrangian of
 chiral boson were constituted
 with the second order time derivative of the field. In the description of Seigel chiral constraint
 was in a quadratic form where as in the description of Srivastava it was in a linear
 form. One more
 ingenious description of chiral boson came from the description of Floreanini
  and Jackiw \cite{FJ}. In this description the lagrangian of chiral boson was constituted
  with first order time derivative of the field. In Ref \cite{JSON}, we find an interesting
  description towards quantization of that free FJ type chiral boson.
  In a very resent work \cite{RPM}, we find an application of augmented super field
  approach to derive the off-shell nilpotent and absolutely anti-commuting
  (anti-)BRST and (anti-)co-BRST symmetry transformations for the
   BRST invariant Lagrangian density of a free chiral boson. Another recent
   important development towards  the BFV quantization of the free chiral boson along with
    study of Hodge decomposition theorem in the context of conserved charges has came in \cite{BPM}

  The obvious generalization of free chiral
  boson is to take into account of the
  interaction of gauge field with that and this interacting field theoretical model is known as gauged
  model of chiral boson. The interacting theory of chiral boson was first described
  by Bellucci, Golterman and Petcher \cite{BEL} with Seigel  like kinetic term
  for chiral boson. So naturally the theory of interacting chiral
  boson with FJ type kinetic was wanted for as free FJ type chiral boson became available in \cite{FJ}
  and that was successfully met up by Harada \cite{KH}.  After
  the work of Harada \cite{KH}, interacting chiral boson based on
  FJ type kinetic term  attracted considerable attention \cite{BAR, BAL, ARC, ABREU, EUNE, SG} in
  spite of the fact that
  this theory of interacting chiral boson was not derived from the iterating theory of chiral boson as
   developed in \cite{BEL}. Harada obtained it from Jackiw-Rajaraman (JR) version of
   chiral Schwinger
   model with an ingenious insertion of a chiral constraint in the
   phase space this theory \cite{JR}. So there is
   a missing link between the two types of interacting gauged chiral
   boson.
An attempt towards search for a link is, therefore, a natural
extension which we would like to explore. In fact, we want to show
whether the gauged model of FJ type chiral boson is contained
within the gauged chiral boson of Seigel type chiral boson which
is available in \cite{BEL}. The study of this model may be
beneficial from another another point of view indeed; where
anomaly is the central issue of investigation \cite{JR, KH, PM,
PMS, ARC, ARPL, SM2, ARUP}, since it is known from Ref. \cite{KH}
that the model took birth from the JR version of chiral Schwinger
model and it is known that  chiral generation of Schwinger model
\cite{SCH} due to Hagen \cite{HAG} gets secured from unitarity
problem when anomaly was taken into consideration in it by Jackiw
and Rajaraman \cite{JR}. In this respect, the recent chiral
generation of Thirring-Wess model is of worth-mentioning
\cite{ARAN1, ARAN2}. So when the issue of searching of the desired
link gets settled down a natural extension that comes
automatically in mind is to study the symmetry underlying in the
model and perform the quantization of the model. BRST quantization
in this context scores over other.

BRST formalism provide a natural framework of covariant
quantization of field theoretical models and is interesting in its
own right since it ensures unitarity and renormalizability of the
theory \cite{BRS1, BRS2, BRS3}. Therefore, BRST quantization of
the gauged chiral boson would certainly be of interest. So we
apply the Batalin, Fradkin and Vilkovisky (BFV) \cite{BAT1, BAT2,
BA3, BA4} formalism in order to get a BRST invariant reformulation
of the said model. In fact, we will use here the improved version
due to FIK \cite{FIK} in our work since it helps to get the
Wess-Zumino \cite{WSJ} term in a transparent way which was found
lacking in the work \cite{SG}. The Wess-Zumino term for the free
chiral boson obtain in \cite{SG}, though agrees with the
conventional Wess-Zumino term that can be inherited from
\cite{KHWES}, the term which was demanded by  the author as the
Wess-Zumino term for the gauged model of chiral boson fails to do
so. Surprisingly, however, the final BRST invariant effective
action for gauged chiral boson presented in \cite{SG} shows on
shell BRST symmetry. So a natural question arises whether or not
FIK formalism fails to produce the appropriate Wess-Zumino term
for the gauged model of FJ type chiral boson since it was found to
be instrumental to get the BRST invariant reformulation for
several physical sensible field theoretical models with the
appropriate Wess-Zumino term \cite{KIM1, KIM2, KIM3, MIAO, PBRA,
AR1, AR2, AR3, AR4}. To explore the above fact, we are in fact,
driven towards the reinvestigation of the BRST invariant
reformulation of the gauged model of FJ type chiral boson.

Gauged model of chiral boson with the  Wess-Zumino term would be a
gauge invariant theory in the extended phase space. So if our
attempt gets a  positive shape towards BRST quantization with the
appearance of appropriate Wess-Zumino term then a natural
extension would be to proceed towards the alternative quantization
of the gauge invariant part of the theory  and the next task would
certainly be to show the equivalence between the physical content
of the actual gauge non-invariant theory and the gauge invariant
theory of the extended phase space which we would also like to
address within this work. Note that this type of investigation is
not possible without the appropriate Wess-Zumino term which was
found lacking in \cite{SG}

The plan of the paper is as follows. In Sec. II we are intended to
find the missing link between the two types  of mutually exclusive
developments of gauged chiral boson.  Sec. III will be devoted
toward the BRST invariant reformulation of the gauged model of
chiral boson which is based on FJ type kinetic term for chiral
boson. In Sec. IV, we quantize the gauge invariant part of the
lagrangian obtained during the process of BRST quantization in
Sec. III with the Lorentz gauge. In Sec. V, an equivalence is made
between the actual gauge non-invariant theory with the gauge
invariant transmuted form obtained in Sec. III.
\section{A gauged model of chiral boson with the Siegel type kinetic term}
The gauged model  of chiral boson with the Siegel type of kinetic
term is described by the lagrangian density \cite{BEL}
    \begin{eqnarray}
     {\cal L}_{B}&=&\frac{1}{2} (\dot{\phi}^{2}-\phi^{\prime 2})+e(\phi^{\prime}+\dot{\phi})(A_{0}-A_{1})
     +\frac{\lambda}{2}[(\dot{\phi}-\phi^{\prime})+e(A_{0}-A_{1})]^{2}\nonumber\\
     &+&\frac{1}{2}(\dot{A_{1}}-A_{0}^{\prime})^{2}+\frac{1}{2}ae^{2}(A_{0}^{2}-A_{1}^{2}).\label{LSCB}
\end{eqnarray}
Here over dot and over prime represent the time and space
derivative respectively. Here $m^2$ is written as $ae^{2}$ for
later convenience. The symbol $e$ indicates the  coupling constant
which has one mass dimension. The momenta corresponding to the
field $A_0$, $A_1$, $\phi$ and $\lambda$ respectively are
\begin{equation}
\frac{\partial {\cal L}_{B}}{\partial \dot{A_{0}}}=\pi_{0}=0
\label{MO1}
\end{equation}
\begin{equation}
\frac{\partial {\cal L}_{B}}{\partial \dot{A_{1}}}=\pi_1 =  \dot
A_1 -A'_0 \label{MO2}
\end{equation}
\begin{equation}
\frac{\partial {\cal
L}_{B}}{\partial\dot{\phi}}=\pi_{\phi}=(1+\lambda)\dot{\phi}
-\lambda\phi^{\prime}+e(1+\lambda)(A_{0}-A_{1})
\label{MO3}
\end{equation}
\begin{equation}
\frac{\partial {\cal L}_{B}}{\partial
\dot{\lambda}}=\pi_{\lambda}=0 \label{MO4}
\end{equation}
The canonical Hamiltonian density of the system is obtained
through a Legendre transformation:
\begin{equation}
{\cal_H}_c=\pi_{\phi}\dot{\phi}+\pi_{1}\dot{A_{1}}-L.
\end{equation}
Using equations (\ref{MO1}),(\ref{MO2}),(\ref{MO3}) and
(\ref{MO4}), we find that ${\cal H}_c$ takes the following form
\begin{eqnarray}
{\cal_H}_c
&=&\frac{\pi_{1}^{2}}{2}+\pi_{1}A_{0}^{\prime}+\pi_{\phi}\phi^{\prime}+\frac{1}{2}e^{2}(A_{1}-A_{0})^{2}
-e(\pi_{\phi}+\phi^{\prime})(A_{0}-A_{1})\nonumber\\
&-& \frac{ae^{2}}{2}(A_{0}^{2}-
A_{1}^{2})+\frac{1}{2(1+\lambda)}{(\pi_{\phi}-\phi^{\prime})}^{2}+u\pi_{0}+v\pi_{\lambda}.
\label{CHAM}
\end{eqnarray}
In equation (\ref{CHAM}), $u$ and $v$ are the two lagrange
multipliers. The following two equations
\begin{equation}
\Omega_1=\pi_{0}\approx 0 ,\label{CON1}
\end{equation}
\begin{equation}
\Omega_2 =\pi_{\lambda}\approx 0, \label{CON2}
\end{equation}
are identified as primary constraints of this system  since these
two do not contain the time derivative of the fields. The
preservation of the constraints (\ref{CON1}) and (\ref{CON2})
leads to the following two constraints:
\begin{equation}
\Omega_3=\pi_{1}^{\prime}+e(\pi_{\phi}+\phi^{\prime})+e^{2}[(a-1)A_{0}
+ A_{1}]\approx 0,\label{CON3}
\end{equation}
\begin{equation}
\Omega_4 =\pi_{\phi}-\phi^{\prime}\approx 0.\label{CON4}
\end{equation}
In order to single out the physical degrees of freedom we proceed
to quantize the theory with the following gauge fixing condition.
\begin{equation}
\Omega_5 =\lambda-f \approx 0\label{CON5}
\end{equation}
The generating functional of this system now reads
\begin{equation}
Z=\int dA_0 dA_1 d\pi_1 d\phi d\pi_\phi d\lambda d\pi_\lambda
e^{i\int d^2x[\pi_{\phi}\dot{\phi}+\pi_{1}\dot{A_{1}}-H]}
\delta(\Omega_1)
\delta(\Omega_2)\delta(\Omega_3)\delta(\Omega_4)\delta(\Omega_5).
\end{equation}
After integrating out of the momenta of the fields we get the
generating functional $Z$ in the following form
\begin{equation}
Z=\int dA_0 dA_1  d\phi  d\lambda  e^{i\int d^2x {\cal L}_{GCB}}
\end{equation}
where
\begin{eqnarray}
{\cal
L}_{GCB}=\dot{\phi}\phi^{\prime}-\phi^{\prime2}+2e\phi^{\prime}(A_{0}-A_{1})-\frac{1}{2}e^{2}(A_{0}-A_{1})^{2}
+\frac{1}{2}ae^{2}(A_{0}^{2}-A_{1}^{2})
+\frac{1}{2}(\dot{A_{1}}-A_{0}^{\prime})^{2}. \label{GCB}
\end{eqnarray}
This is the gauged model of chiral boson with FJ type kinetic
term. Note that ${\cal L}_{GCB}$ represents a lagrangian density
that has generated from ${\cal L}_{B}$ and it agrees with the
lagrangian found in \cite{KH}. So we find that the gauged model of
chiral boson with FJ type kinetic term is contained within the
gauged version of Siegel like chiral boson \cite{BEL}.

It is beneficial  to compute the Dirac brackets for completeness
of the analysis since it is a constrained theory and ordinary
Poisson brackets become inadequate for the theories endowed with
constraint. The Dirac bracket \cite{DIR} for the two field
variables $A$ and $B$ is defined by
 \begin{equation}
 [A(x), B(y)]^* = [A(x), B(y)] - \int[A(x) \omega_i(\eta)]
 C^{-1}_{ij}(\eta, z)[\omega_j(z), B(y)]d\eta dz, \label{DDB}
 \end{equation}
 where $C^{-1}_{ij}(x,y)$ is defined by
 \begin{equation}
 \int C^{-1}_{ij}(x,z) [\Omega_i(z), \Omega_j(y)]dz = 1.
 \label{IMAT} \end{equation}
 Here $\Omega_i$'s stands for the standing second class constraints embedded in the phase space of
 the theory.
Therefore, to compute Dirac brackets we need to construct the
matrix constituted with the Poison brackets of the constraints
(\ref{CON1}), (\ref{CON2}),(\ref{CON3}),(\ref{CON4}) and
(\ref{CON5}). The required matrix is
\begin{eqnarray}
C_{ij}=\left(\begin{array}{ccccc}
0 & 0 & -e^{2}(a-1) & 0&0 \\
 0  & 0 & 0  & 0&-1 \\
       e^{2}(a-1) & 0  & 0&0 & 0 \\
 0 &0  &0  &-2\partial &0\\
      0&1 &0 &0 &0
\end{array}\right)\delta(x-y). \label{altere15}
\end{eqnarray}
The matrix $C_{ij}$ is nonsingular. So inverse of it exists which
is found out to be
\begin{eqnarray}
C_{ij}^{-1}=\left(\begin{array}{ccccc}

    0&0&\frac{1}{e^{2}(a-1)} \delta(x-y) & 0  &0 \\
    0 & 0 & 0 &0 & 1 \\
      -\frac{1}{e^{2}(a-1)}\delta(x-y) & 0  & 0&0& 0 \\
       0& 0 & 0& -\frac{1}{4}\epsilon(x-y)& 0\\
            0 & -1 & 0 & 0 & 0\\
\end{array}\right). \label{altere17}
\end{eqnarray}
Here $\epsilon(x)$ is the sign function,$\epsilon(x) =+1$ for $x>
0$  and $\epsilon(x) =-1$ for $x< 0$. $\frac{d}{dx}\epsilon(x) =
2\delta(x)$ Using the definition (\ref{DDB}), straightforward
calculations renders the following Dirac brackets between the
field variables.
\begin{equation}
[A_{0}(x),A_{1}(y)]^{*}=\frac{1}{e^{2}(a-1)}\delta'(x-y),
\label{DB1}
\end{equation}
\begin{equation}
  [\phi(x),\phi(y)]^{*}=-\frac{1}{4}\epsilon{(x-y)},\label{DB2}
  \end{equation}
\begin{equation}
[A_{0}(x),\phi(y)]^{*}=\frac{1}{e(a-1)}\delta(x-y), \label{DB3}
 \end{equation}
\begin{equation}
[A_{0}(x),\pi_{1}(y)]^{*}=-\frac{1}{(a-1)}\delta(x-y), \label{DB4}
\end{equation}
\begin{equation}
 [A_{0}(x),\pi_{\phi}(y)]^{*}=-\frac{1}{e(a-1)}\delta'(x-y), \label{DB5}
 \end{equation}
 \begin{equation}
  [A_{1}(x),\pi_{1}(y)]^{*}=\delta{(x-y)}, \label{DB6}
  \end{equation}
   \begin{equation}
  [\phi(x),\pi_{\phi}(y)]^{*}=\delta{(x-y)}, \label{DB7}
  \end{equation}
Here $(*)$ indicate the Dirac bracket. Here we end up the
description of this Sec. and in the following section we proceed
to wards BRST quantization.
\section{BRST quantization of the gauged model of chiral boson with
FJ type kinetic term} In this section we are intendant to carry
out the BRST quantization of the gauged model of chiral boson with
FJ type kinetic term using the BFV based improved version of FIK
since we are familiar with the several successful attempts with
this improved version towards the generation of the appropriate
Wess-Zumino term during the process of BRST quantization
\cite{KIM1, KIM2, KIM3, MIAO, PBRA, AR1, AR2, AR3, AR4}.

According to this formalism $H_{m}$  is usually known as the
minimal Hamiltonian which is defined by
\begin{equation}
H_m= H_c + \bar{P}_aV^a_bC^b, \label{IM}
\end{equation}
and the  BRST charge Q and  the gauge fixing function G  have the
following expressions respectively:
\begin{equation}
Q=C^a\omega_a-\frac{1}{2}C^bC^cU^a_{cb}\bar{p}_a + P^a\pi_a,
\end{equation}
\begin{equation}
G=\bar{C}_a\chi^a + \bar{P}_a\lambda^a.
\end{equation}
The structure coefficients $U^{c}_{ab}$ and $V^{a}_{b}$ come from
the Poisson brackets among the constraints $\Omega$'s themselves
of the theory and the Poisson's brackets with the canonical
Hamiltonian $H_{c}(q^{i},p_{i})$:
\begin{equation}
[\Omega_{a},\Omega_{b}]=i\Omega_{c}U^{c}_{ab},
[H_{c},\Omega_{a}]=iw_{b}V^{a}_{b},
\end{equation}
where $w_{a}$ (a= 1,2..........N) represents the $N$ number of
constraints embedded in the phase space of the theory defined by
the Hamiltonian $H_{c}(q^{i},p_{i})$. In order to single out the
physical degrees  of freedom $N$ number of additional conditions
$\phi^{a}=0$, are are required to impose within the phase space.
The constraints $\phi^{a}=0$ and $\Omega_{a}=0$ together with the
Hamiltonian equations may be obtained from the action
\begin{equation}
S=\int{[p_{i}\dot{q^{i}}-H(p_{i},q^{i})-\lambda^{a}\Omega_{a}+\pi_{a}\phi^{a}]}dt,
\end{equation}
where $\lambda^{a}$  and $\pi_{a}$ are lagrange multiplier having
Poisson's bracket
 $[\lambda^{a},\pi^{a}]=i\delta^{a}_{b}$ and  $\lambda^{a}$ is contained within the gauge fixing
conditions in the form $\phi^a=\lambda^{a}+ \chi^a$. The variables
$ (C^{i} \bar{P_{i}})$ and $(P^{i} ,\bar{C_{i}})$  are the two
sets of canonically conjugate anti-commuting ghost coordinates and
momenta having the algebra $[C_{i},\bar{P_{i}}]=i\delta(x-y)$ and
$[P^{i}, \bar{C_{i}}]=i\delta(x-y)$. The lagrange multiplier
$\lambda^{a}$  and $\pi_{a}$  have the Poisson bracket
 $[\lambda^{a},\pi^{a}]=i\delta^{a}_{b}$.  The quantum theory, therefore, can be described by the
generating functional
\begin{equation}
Z_G=\int dq^i dp_1 d\lambda^a d\pi_a dC^a d\bar{P}_a dP^a
d\bar{C}_a e^{iS_G},
\end{equation}
where the action $S_G$ is
\begin{equation}
S_G=\int[p_{i}\dot{q^{i}}+\bar{P^{i}}C_{i}+\bar{C^{a}}\dot{P_{a}}-H_{m}+\dot{\lambda^{a}}\pi_{a}
+i[Q,G]]dt.\label{BM}
\end{equation}
With this input, we may proceed towards the BRST quantization of
theory under consideration. The lagrangian density of gauged model
of FJ type chiral boson is given by
\begin{equation}
{\cal
L}_{CB}=\dot{\phi}\phi^{\prime}-\phi^{\prime2}+2e\phi^{\prime}(A_{0}-A_{1})-\frac{1}{2}e^{2}(A_{0}-A_{1})^{2}
+\frac{1}{2}ae^{2}(A_{0}^{2}-A_{1}^{2})+\frac{1}{2}(\dot{A_{1}}-A_{0}^{\prime})^{2}.
\label{FJC}
\end{equation}
For  this  lagrangian density (\ref{FJC}) the canonical momenta
corresponding to the field $A_0$, $A_1$ and $\phi$ respectively
are
\begin{equation}
\frac{\partial {\cal L}_{CB}}{\partial
\dot{A_{0}}}=\pi_{0}=0,\label{MG1}
\end{equation}
\begin{equation}
\frac{\partial {\cal L}_{CB}}{\partial
\dot{A_{1}}}=\pi_{1}=\dot{A_{1}}-A_{0}^{\prime}, \label{MG2}
\end{equation}
\begin{equation}
\frac{\partial {\cal L}_{CB}}{\partial
\dot{\phi}}=\pi_{\phi}=\phi^{\prime}.\label{MG3}
\end{equation}
The canonical hamiltonian can be calculated using equations
(\ref{MG1}),(\ref{MG2}) and (\ref{MG3}) through a Legendre
transformation as has been done earlier:
\begin{equation}
H_{c}=\int dx[\frac{1}{2}\pi_{1}^{2}+\pi_{1}A_{0}^{\prime} +
\phi^{\prime2}+ -2e\phi^{\prime}(A_{0}-A_{1})
+\frac{1}{2}e^{2}(A_{0}-A_{1})^{2}
-\frac{1}{2}ae^{2}(A_{0}^{2}-A_{1}^{2})]\label{CHAMF}
\end{equation}
Equation (\ref{MG1}) and (\ref{MG3}), do not contain any time
derivative of the fields. So these two are the primary constraint
of the theory.
\begin{equation}
\omega_{1}=\pi_{\phi}-\phi^{\prime} \approx 0, \label{CONPI}
\end{equation}
\begin{equation}
\omega_{2}=\pi_{0}\approx 0. \label{CONCHI}
\end{equation}
Therefore, the Hamiltonian  reads
\begin{eqnarray}
H_{P}=\int dx[\frac{1}{2}\pi_{1}^{2}+\pi_{1}A_{0}^{\prime}
+\phi^{\prime2}
2e\phi^{\prime}(A_{0}-A_{1})+\frac{1}{2}e^{2}(A_{0}-A_{1})^{2}
-\frac{1}{2}ae^{2}(A_{0}^{2}-A_{1}^{2})+u(\pi_{\phi}-\phi^{\prime})+v\pi_{0}].\label{EHAMF}
\end{eqnarray}
 The preservation  of $\omega_{2}$ renders the following new
 constraint
\begin{equation}
\omega_{3}=\pi_{1}^{\prime}+2e\phi^{\prime}+e^{2}(a-1)A_{0}+e^{2}A_{1}\approx
\pi_{1}^{\prime}+e\phi^{\prime}+e\pi_{\phi}+e^{2}(a-1)A_{0}+e^{2}A_{1}\approx0.
\label{CONGA}
\end{equation}
  The preservations of $\omega_{1}$ and
$\omega_{3}$ however do not give rise to any new constraint. These
two conditions fix the velocities $u$ and $v$ respectively:
\begin{equation} u=A_{1}^{\prime}-\frac{1}{(a-1)}\pi_{1}
\label{VELV},
\end{equation}
and
\begin{equation}
v=\phi^{\prime}-e(A_{0}-A_{1}). \label{VELU}
\end{equation}
Note that the constraints labelled by $\Omega's$ in Sec. II
differs in number with the constraints labelled by $\omega's$ in
this section because of the presence lagrange multiplier $\lambda$
in that section. However careful look reveals that
$\Omega_1\approx\omega_2, \Omega_3\approx\omega_3$ and
$\Omega_4\approx\omega_1$. Now imposing the expression of $u$ and
$v$ in (\ref{EHAMF}) the Hamiltonian turns into
\begin{eqnarray}
H_{P}&=&\int dx
[\frac{1}{2}\pi_{1}^{2}+\pi_{1}A_{0}^{\prime}+\pi_{\phi}(\phi^{\prime}-e(A_{0}-A_{1}))
-e\phi^{\prime}(A_{0}-A_{1})
+\frac{1}{2}e^{2}(A_{0}-A_{1})^{2}\nonumber\\
&-&\frac{1}{2}ae^{2}(A_{0}^{2}-A_{1}^{2})
+\pi_{0}(A_{1}^{\prime}-\frac{1}{(a-1)}\pi_{1})+
(\pi_{\phi}-\phi^{\prime})(\phi^{\prime}-e(A_{0}-A_{1}))]\label{RHAM}
\end{eqnarray}
The constraints of the theory satisfy the following Poisson
brackets between themselves
\begin{equation}
[\omega_{1},\omega_{1}]=-2i\delta'(x-y), \label{CPB1}
\end{equation}
\begin{equation}
[\omega_{1},\omega_{3}] =0,\label{CPB2}
\end{equation}
\begin{equation}
[\omega_{2},\omega_{2}] = 0.\label{CPB3}
\end{equation}
\begin{equation}
[\omega_{2},\omega_{3}]=-ie^{2}(a-1)\delta(x-y).\label{CPB4}
\end{equation}
The involution relation between the Hamiltonian and  the
constraints $\omega_{1}$, $\omega_{2}$ and $\omega_{3}$ are
\begin{equation}
-i[\omega_{1},H_{P}]=\omega_{1}^{\prime}, \label{TEV1}
\end{equation}
\begin{equation}
-i[\omega_{2},H_{P}]=\omega_{3}, \label{TEV2}
\end{equation}
\begin{equation}
-i[\omega_{3},H_{P}]=\omega_{2}^{\prime\prime}-\frac{e^{2}}{(a-1)}\omega_{2}.
\label{TEV3}
\end{equation}
The set of second class constraints  $\omega_{1}$, $\omega_{2}$
and $\omega_{3}$  can be converted into a first a class set with
the help of two  auxiliary canonical pairs $(\theta, \pi_\theta)$
and $(\eta, \pi_\eta)$. The first class set of constraints those
which are constructed from the said second class set of
constraints using these auxiliary fields  are the following.
\begin{equation}
\bar{\omega_{1}}=\omega_{1}+\pi_{\theta}+\theta^{\prime},
\label{FCON1}
\end{equation}
\begin{equation}
\bar{\omega_{2}}=\omega_{2}-\pi_{\eta}, \label{FCON2}
\end{equation}
\begin{equation}
\bar{\omega_{3}}=\omega_{3}+e^{2}(a-1)\eta. \label{FCON3}
\end{equation}
The Hamiltonian consistent with the first class set of constraint
({\ref{FCON1}), ({\ref{FCON2}) and ({\ref{FCON3}) is
\begin{equation}
\bar{H}=H_{P}+H_{BF}, \label{FCHAM}
\end{equation}
where $H_{BF}$ would certainly be constituted with the auxiliary
fields which is found out to be
\begin{equation}
H_{BF}=\int dx[
\frac{1}{4}(\pi_{\theta}+\theta^{\prime})^{2}+\frac{1}{2}e^{2}(a-1)\eta^{2}+\frac{1}{2e^{2}(a-1)}\pi_{\eta}^{\prime
2}+\frac{1}{2(a-1)^{2}}\pi_{\eta}^{2}] \label{fg16}
\end{equation}
For consistency, the time evaluation of these first class set must
be identical to the (\ref{TEV1}),(\ref{TEV2}) and (\ref{TEV3}).
Precisely these are the following.
\begin{equation}
-i[\bar{\omega_{1}},\bar{H}]=\bar{\omega_{1}^{\prime}},
\label{fg18}
\end{equation}
\begin{equation}
-i[\bar{\omega_{2}},\bar{H}]=\bar{\omega_{3}}, \label{fg19}
\end{equation}
\begin{equation}
-i[\bar{\omega_{3}},\bar{H}]=\bar{\omega_{2}}^{\prime\prime}-\frac{e^{2}}{(a-1)}\bar{\omega_{2}}.
\label{fg20}
\end{equation}
The stage is now set to introduce the two pairs of ghost and
anti-ghost fields $(C^{i} ,\bar{P_{i}})$ and $(P^{i}
,\bar{C_{i}})$.
 We also need to introduce a pair of multiplier fields $(N_{i}, B_{i})$
The multipliers and the ghost anti-ghost pairs  satisfy the
following canonical Poisson's Brackets: $[C^{i} ,\bar{P_{j}}]=
[P^{i} ,\bar{C_{j}}]= [N^{i},B_{j}]=i\delta^{i}_{j}\delta(x-y)$,
where $i=1,2,3$. According to the definition
\begin{equation}
H_{BRST}=H_{P}+H_{BF}+\int dx \bar{P_{a}}V^{a}_{b}C^{b},
\label{HBRST}
\end{equation}
and
\begin{equation}
H_{U}=H_{BRST}+i\int dx [Q,G].\label{HU}
\end{equation}
In this situation BRST charge Q and the fermionic gauge fixing
function G can be written down as
\begin{equation}
Q=\int{dx(C^{i}\bar{\omega_{i}}+P^{i}B_{i})},\label{BRSC}
\end{equation}
\begin{equation}
G=\int{dx(\bar{C_{i}}\chi^{i} +\bar{P_{i}}N^{i})}. \label{GFIX}
\end{equation}
We are now in a position to fix up the gauge condition which is
very crucial for achieving appropriate Wess-Zumino term. It is
found that the following gauge fixing conditions render the
required service to wards that end.
\begin{equation}
\chi_{1}=\pi_{\phi}-\phi^{\prime}, \label{GFIX1}
\end{equation}
\begin{equation}
\chi_{2}=\dot{N^{2}}+A_{0}, \label{GFIX2}
\end{equation}
\begin{equation}
\chi_{3}=-A_{1}^{\prime}+\frac{\alpha}{2}B_{3}.\label{GFIX3}
\end{equation}
Let us now calculate the commutation relation between the BRST
charge and the gauge fixing function:
\begin{equation}
[Q,G]=B_{i}\chi^{i}+P_{i}P^{i}-C^{3}\bar{C_{3}}^{\prime\prime}+
\bar{\omega_{i}}N^{i}-C^{2}\bar{C_{2}}-2C^{1}\bar{C_{1}}.
\end{equation}
Generating functional for this system can be written down as
\begin{equation}
Z =\int[D\mu]\exp^{iS}.
\end{equation}
where $[D\mu]$ is the Liouville measure in the extended phase
space.
\begin{equation}
d\mu=[d\phi][d\pi_{\phi}] \sum_{i=0}^{1}[dA_{i}][d\pi_{i}]
[d\eta][d\pi_{\eta}]
[d\theta][d\pi_{\theta}]\sum_{k=1}^{3}[dN^{k}][dB_{k}][dC^{k}],
[d\bar{C_{k}}][d P^{k}],[d\bar{P_{k}}], \label{LAG34}
\end{equation}
and the action $S$ is explicitly given  by
\begin{equation}
 S=\int d^{2}x[\dot{\phi}\pi_{\phi}+\dot{A_{1}}\pi_{1}
+\dot{A_{0}}\pi_{0}+\dot{\eta}\pi_{\eta} +\dot{\theta}\pi_{\theta}
+\dot{N_{i}}B_{i} +\bar{P_{i}}\dot{C^{i}}
+\bar{C_{i}}\dot{P^{i}}-H_{U}].
\end{equation}
The above formulation allows the following simplification:
\begin{equation}
\int dx(B_{i} N^{i} + \bar{C_{i}} \dot{P^{i}}) = −i[Q, \int
dx\bar{C_{i}}\dot{N^{i}}]. \label{LEG}
\end{equation}
Exploiting the above simplification (\ref{LEG}) we obtain the
effective action in the following form.
\begin{eqnarray}
S_{eff}&=&\int d^{2}x[\dot{\phi}\pi_{\phi}+\dot{A_{1}}\pi_{1}
+\dot{A_{0}}\pi_{0}+\dot{\eta}\pi_{\eta} +\dot{\theta}\pi_{\theta}
+\dot{N^{2}}B_{2}
+\dot{N^{3}}B_{3}+\bar{P_{1}}\dot{C^{1}}+\bar{P_{2}}\dot{C^{2}}
\nonumber\\
&+&\bar{P_{3}}\dot{C^{3}}+\bar{C_{2}}\dot{P^{2}}+\bar{C_{3}}\dot{P^{3}}-P_{1}\bar{P^{1}}
-P_{2}\bar{P^{2}}-P_{3}\bar{P^{3}}
-[\pi_{\phi}(\phi^{\prime}
- e(A_{0}-A_{1}))\nonumber\\
&+&\frac{1}{2}\pi_{1}^{2}+\pi_{1}A_{0}^{\prime}-e\phi^{\prime}(A_{0}-A_{1})+\frac{1}{2}e^{2}(A_{0}-A_{1})^{2}
-\frac{1}{2}ae^{2}(A_{0}^{2}-A_{1}^{2})\nonumber\\
&+&\pi_{0}(A_{1}^{\prime}-\frac{1}{(a-1)}\pi_{1})
+\frac{1}{4}(\pi_{\theta}+\theta^{\prime})^{2}
+\frac{1}{2}e^{2}(a-1)\eta^{2}+\frac{1}{2e^{2}(a-1)}\pi_{\eta}^{\prime
2} +\frac{1}{2(a-1)^{2}}\pi_{\eta}^{2}]\nonumber\\
&+&(\pi_{\phi}-\phi^{\prime}+\pi_{\theta}+\theta^{\prime})N^{1}
+(\pi_{0}-\pi_{\eta})N^{2}+(\pi_{1}^{\prime}
+e\phi^{\prime}+e\pi_{\phi}+e^{2}(a-1)A_{0}+e^{2}A_{1}+e^{2}(a-1)\eta)N^{3}\nonumber\\
&-&B_{1}\chi^{1}-B_{2}\chi^{2}-B_{3}\chi^{3}
-\bar{P_{1}}C_{1}^{\prime}+\bar{P_{3}}C_{2}
-\bar{P_{2}^{\prime\prime}}C_{3}-\frac{1}{(a-1)}e^{2}\bar{P_{2}}C_{3}
-C^{2}\bar{C_{2}}+C^{3}\bar{C_{3}}^{\prime\prime}
+2C^{1}\bar{C_{1}}]. \label{SEFF}
\end{eqnarray}
We are in a state to integrate out of the fields
$\pi_1$,$\pi_0$,$\eta$,$B_1$,$B_2$,$N^1$,$N^2$
$\bar{C^{1}}$,$\bar{P_{1}}$,$\bar{P_{3}}$,$\bar{P_{2}}$, one by
one to have the  action in a desired shape. After integrating out
of the said fields and choosing $N_{3}=A_{0}$, the action reduces
to
\begin{eqnarray}
S_{eff}&=&\int
d^{2}x[\dot{\phi}\phi^{\prime}-\phi^{\prime2}+2e\phi^{\prime}(A_{0}-A_{1})
-\frac{1}{2}e^{2}(A_{0}-A_{1})^{2}+\frac{1}{2}ae^{2}(A_{0}-A_{1})
+\frac{1}{2}(\dot{A_{1}}-A_{0}^{\prime})^{2}+\frac{1}{(a-1)}\pi_{\eta}(\dot{A_{1}}-A_{0}^{\prime})\nonumber\\
&+&(\pi_{\eta}^{\prime}A_{1} -\dot{\pi_{\eta}}A_{0})
+\frac{1}{2e^{2}(a-1)}(\dot{\pi_{\eta}}^{2}-\pi_{\eta}^{\prime 2})
+B_{3}\dot{A_{0}}-B_{3}{A_{1}}^{\prime}+\frac{\alpha}{2}B_{3}^{2}+\partial_{\mu}C^{3}\partial^{\mu}\bar{C_{3}}]
\label{fg33}.
\end{eqnarray}
If we  now define $\pi_{\eta}=e(a-1)\eta$, $C_{3}=C$, and
$B_{3}=B$ we get the desired BRST invariant effective action:
\begin{eqnarray}
S_{BRST}&=&\int
d^{2}x[\dot{\phi}\phi^{\prime}-\phi^{\prime2}+2e\phi^{\prime}(A_{0}-A_{1})-\frac{1}{2}e^{2}(A_{0}-A_{1})^{2}
+\frac{1}{2}ae^{2}(A_{0}^{2}-A_{1}^{2})
+\frac{1}{2}(\dot{A_{1}}-A_{0}^{\prime})^{2}+\frac{1}{2}(a-1)(\dot{\eta}^{2}-\eta^{\prime 2})\nonumber\\
&+& e(A_{0}\eta^{\prime}-A_{1}\dot{\eta}) +
e(a-1)(A_{1}\eta^{\prime}-A_{0}\dot{\eta})+\partial_{\mu}C\partial^{\mu}\bar{C}
+ B\partial_{\mu}A^{\mu}+\frac{\alpha B^{2}}{2}]. \label{EABR}
\end{eqnarray}
The  action (\ref{EABR}) is now found to remain invariant if the
fields transform as follows.
\begin{equation}
\delta{\phi}=e\lambda C, \label{fg36}
\end{equation}
\begin{equation}
 \delta A_{0}=-\lambda\dot{C}, \label{fg37}
 \end{equation}
\begin{equation}
\delta{A_{1}}=-\lambda C^{\prime}, \label{fg38}
\end{equation}
\begin{equation}
\delta{\eta}=-\lambda eC, \label{fg39}
\end{equation}
\begin{equation}
\delta C=0,  \label{fg45}
\end{equation}
\begin{equation}
 \delta \bar{C}=-\lambda B.
 \end{equation}
The above transformations are the very BRST transformation
generated from the BRST charge (\ref{BRSC}). The Wess-Zumino term
for the theory under consideration can easily be identified as
\begin{equation}
S_{WESS}=\int d^2x [\frac{1}{2}(a-1)(\dot{\eta}^{2}-\eta^{\prime
2})+ e(A_{0}\eta^{\prime}-A_{1}\dot{\eta}) +
e(a-1)(A_{1}\eta^{\prime}-A_{0}\dot{\eta})]. \label{WESZ}
\end{equation}
This very action (\ref{WESZ}) contains the appropriate Wess-Zumino
term corresponding to the theory of our present consideration and
it agrees with the Ref. \cite{KHWES}. We would like to reiterate
that in \cite{SG} it was lacking for. In fact, in \cite{SG}, the
term which was demanded  by the author as the Wess-Zumino term
though do not agree with Ref. \cite{KHWES}, nevertheless,  the
author finds on shell BRST invariance with that Wess-Zumino term.
The term standing in equation (\ref{WESZ}) however establishes the
off-shell BRST invariance. To achieve the appropriate Wess-Zumino
term for this theory which agrees with \cite{KHWES},  is a novel
aspect of this reinvestigation.

\section{{\bf An  alternative quantization of the gauge invariant version of the theory}}
The quantization of gauged model of FJ type chiral bosom was
available in \cite{KH}. It was attempted there to quantize it in a
gauge non-invariant manner. The gauge invariant version certainly
can be quantized. We refer the works \cite{SM1, AR4}, where the
authors made alternative quantization of chiral Schwinger model
with the Faddeevian anomaly and generalized version of QED where
vector and axial vector interaction gets mixed up with different
wight respectively. Some gauge fixing is needed in this situation
indeed. We choose the Lorentz gauge and proceed to quantize the
gauge symmetric version of the gauged model of FJ chiral boson.
The gauge symmetric version of the said theory with Lorentz gauge
is described by the lagrangian density.
\begin{eqnarray}
{\cal
L}&=&\dot{\phi}\phi^{\prime}-\phi^{\prime2}+2e\phi^{\prime}(A_{0}-A_{1})-\frac{1}{2}e^{2}(A_{0}-A_{1})^{2}
+\frac{1}{2}ae^{2}(A_{0}^{2}-A_{1}^{2})
+\frac{1}{2}(\dot{A_{1}}-A_{0}^{\prime})^{2}+\frac{1}{2}(a-1)(\dot{\eta}^{2}-\eta^{\prime 2})\nonumber\\
&+& e(A_{0}\eta^{\prime}-A_{1}\dot{\eta}) +
e(a-1)(A_{1}\eta^{\prime}-A_{0}\dot{\eta})+
B\partial_{\mu}A^{\mu}+\frac{\alpha B^{2}}{2}. \label{GFCB}
\end{eqnarray}
Gauge fixing is needed in order to single out the real physical
degrees of freedom from the gauge symmetric version of the
extended phase space. The Euler-Lagrange equations of motion
corresponding to the fields $\phi$, $A_0$, $A_1$, $B$ and $\eta$
that follow from the lagrangian density (\ref{GFCB}) respectively
are
\begin{equation}
\dot{\phi^{\prime}}-\phi^{\prime\prime}+e(A_{0}^{\prime}-A_{1}^{\prime})=0,
\label{FCE1}
\end{equation}
\begin{equation}
A_{0}^{\prime\prime}-\dot{A_{1}}^{\prime}+e^{2}(1-a)A_{0}-e^{2}A_{1}+e(a-1)\dot{\eta}-e\eta^{\prime}
-2e\phi^{\prime}-\dot{B}=0,\label{FCE2}
\end{equation}
\begin{equation}
\ddot{A_{1}}-\dot{A_{0}}^{\prime}+ae^{2}A_{1}+e^{2}A_{1}-e^{2}A_{0}-e(a-1){\eta^{\prime}}
+e\dot{\eta}+2e\phi^{\prime}+B^{\prime}=0, \label{FCE3}
\end{equation}
\begin{equation}
\partial_{\mu}A^{\mu}+\alpha B=0, \label{FCE4}
\end{equation}
\begin{equation}
(a-1)\ddot{\eta}-(a-1)\eta^{\prime\prime}-e(a-1)\dot{A_{0}}+e(a-1)A_{1}^{\prime}+eA_{0}^{\prime}-e\dot{A_{1}}=0.
\label{FCE5}
\end{equation}
It is found that the following expression of $A_\mu$, $\phi$ and
$\eta$ represents the exact solution of the equations
(\ref{FCE1}),(\ref{FCE2}),(\ref{FCE3}),(\ref{FCE4}) and
(\ref{FCE5})
\begin{equation}
A_{\mu}=\frac{1}{ae^{2}}[-\frac{(a-1)}{a}\tilde{\partial_{\mu}}F+\partial_{\mu}{B}
-e\tilde{\partial_{\mu}}h-ea\partial_{\mu}\zeta],\label{FCE6}
\end{equation}
\begin{equation}
\phi=-\frac{(a-1)}{ea^{2}}F-\frac{h}{a}+\zeta, \label{fg672}
\end{equation}
\begin{equation}
\eta=-\frac{F}{ea^{2}}-\zeta-\frac{h}{a(a-1)}. \label{fg673}
\end{equation}
if the following free field equations
\begin{equation}
(\partial_{0}-\partial_{1}){h}=0, \label{CD1}
\end{equation}
\begin{equation}
 (\partial_{0}-\partial_{1})B=0, \label{CD2}
\end{equation}
\begin{equation}
\Box\zeta=\alpha e B,\label{CD3}
\end{equation}
\begin{equation}
[\Box+m^{2}]F=0, \label{CD4}
\end{equation}
\begin{equation}
 m^{2}=\frac{a^{2}e^{2}}{(a-1)},\label{CD5}
\end{equation}
are maintained. Therefore, the free fields in terms of which the
system is completely described are
\begin{equation}
h=-(a-1)(\phi+\eta+\frac{1}{ea}F), \label{FF1}
\end{equation}
\begin{equation}
\zeta=\frac{1}{a}\phi-\frac{(a-1)}{a}\eta,\label{FF2}
\end{equation}
\begin{equation}
F=\pi_{1},   \label{FF3}
\end{equation}
\begin{equation}
B=\pi_{0}.  \label{FF4}
\end{equation}
The equal time commutation relations corresponding to the free
fields are found out to be
\begin{equation}
[F, \dot{F}]= im^2\delta(x-y),\label{CFF0}
\end{equation}
\begin{equation}
 [\zeta,\dot{\zeta}]=i\frac{1}{a}\delta(x-y), \label{CFF1}
\end{equation}
\begin{equation}
[h,\dot{h}]=i\delta(x-y), \label{CFF2}
\end{equation}
\begin{equation}
[B,\dot{\zeta}]=ie\delta(x-y).\label{CFF3}
\end{equation}
Note that $F=\pi_1$ represents a massive field with mass $m$ and
$h$ represents a massless chiral boson. These two are the replica
of the spectrum as obtained in \cite{KH}. The equations involving
$B$ appear because of the presence of the auxiliary field in the
Lorentz gauge fixing. Note that B has the vanishing commutation
relation with the physical field $B$ and $h$. The field $\zeta$
represents the zero mass dipole field playing the role of gauge
degrees of freedom that can be eliminated  by operator gauge
transformation. So the theoretical spectrum agrees in an exact
manner with the theoretical spectrum obtained in \cite{KH}

\section{To Show the equivalence  between the gauge invariant and gauge variant version of the model}
In this section an attempt is made to show the equivalence between
the gauge invariant version of the extended phase space and the
gauge variant version of the usual phase space of the gauged model
of FJ chiral boson. It is important because to make the model
gauge invariant phase space was needed to extend introducing the
Wess-Zumino fields. So, what service does the Wess-Zumino fields
actually renders is a matter of utter curiosity.

To meet it let us start with the lagrangian of the gauged FJ type
chiral boson with the appropriate Wess-Zumino term as is obtained
from our investigation. The said lagrangian density reads
\begin{eqnarray}
{\cal
L}&=&\dot{\phi}\phi^{\prime}-\phi^{\prime2}+2e\phi^{\prime}(A_{0}-A_{1})
-\frac{1}{2}e^{2}(A_{0}-A_{1})^{2}+\frac{1}{2}ae^{2}(A_{0}^{2}-A_{1}^{2})\nonumber\\
&+&\frac{1}{2}(\dot{A_{1}}-A_{0}^{\prime})^{2}+\frac{1}{2}(a-1)(\dot{\eta}^{2}
-\eta^{\prime 2})\nonumber\\
&+& e(A_{0}\eta^{\prime}-A_{1}\dot{\eta}) +
e(a-1)(A_{1}\eta^{\prime}-A_{0}\dot{\eta}).\label{ALTER}
\end{eqnarray}
To show the equivalence between the gauge invariant and the gauge
variant version of this model we proceed with computation of the
canonical momenta corresponding to the fields $\phi$, $A_0$, $A_1$
and $\eta$:
\begin{equation}
\frac{\partial L}{\partial{\dot{\phi}}}=\pi_{\phi}=\phi^{\prime},
\label{ALM1}
\end{equation}
\begin{equation}
\frac{\partial L}{\partial{\dot{A_{0}}}}=\pi_{0}=0, \label{ALM2}
\end{equation}
\begin{equation}
\frac{\partial
L}{\partial\dot{A_{1}}}=\pi_{1}=\dot{A_{1}}-A_{0}^{\prime}.\label{ALM3}
\end{equation}
\begin{equation}
\frac{\partial L}{\partial\dot{\eta}}=\pi_{\eta}
=(a-1)\dot{\eta}-eA_{1}-e(a-1)A_{0}.\label{ALM4}
\end{equation}
The equations (\ref{ALM1}) and (\ref{ALM2}) are independent of
velocity so these two represent the two primary constraints.
Explicitly these two are
\begin{equation}
T_{1}=\pi_{0}\approx 0,
\end{equation}
\begin{equation}
T_{2}=\pi_{\phi}-\phi^{\prime}\approx 0.
\end{equation}
Using the equations (\ref{ALM1}),(\ref{ALM2}),(\ref{ALM3}) and
(\ref{ALM4}), a Legendre transformation leads to the canonical
Hamiltonian $H_{c}$ corresponding to the lagrangian density
(\ref{ALTER}):
\begin{eqnarray}
H_{c}&=&\int dx[
\frac{1}{2}\pi_{1}^{2}+\pi_{1}A_{0}^{\prime}+\phi^{\prime2}-2e\phi^{\prime}(A_{0}-A_{1})
+\frac{1}{2}e^{2}(A_{0}-A_{1})^{2}\nonumber\\
&-&\frac{1}{2}ae^{2}(A_{0}^{2}-A_{1}^{2 })
+\frac{1}{2}(a-1)\eta^{\prime2}-eA_{0}\eta^{\prime}-e(a-1)A_{1}\eta^{\prime}\nonumber\\
&+&\frac{1}{2(a-1)}\pi_{\eta}^{2}+\frac{e^{2}}{2(a-1)}((a-1)A_{0}+A_{1})^{2})\nonumber\\
&+&\frac{e}{(a-1)}\pi_{\eta}((a-1)A_{0}+eA_{1})].\label{HALTER}
\end{eqnarray}
The preservation of the constraint of $T_{1}$ leads to a new
constraint
\begin{equation}
T_{3}=\pi_{1}^{\prime}+e\pi_{\phi}+e\phi^{\prime}+e\eta^{\prime}-e\pi_{\eta}\approx
0 . \label{alter11}
\end{equation}
The ref. \cite{FALCK}, suggests that we have to choose appropriate
gauge fixing at this stage to meet our need and we find that gauge
fixing conditions those which have been found suitable for this
system are the following:
\begin{equation}
T_{4}=e\eta^{\prime}\approx 0,\label{ALGF1}
\end{equation}
\begin{equation}
T_{5}=e^{2}(a-1)A_{0}+e^{2}A_{1}+e\pi_{\eta}\approx 0
.\label{ALGF2}
\end{equation}
Under insertion of the conditions of (\ref{ALGF1}) and
(\ref{ALGF2}),  $T_{3}$ and $H_{c}$, turns into $\tilde{T}_{3}$
and $\tilde{H}_{c}$  those which are explicitly given by
\begin{equation}
\tilde{T}_{3}=\pi_{1}^{\prime}+e\pi_{\phi}+e\phi^{\prime}+e^{2}(a-1)A_{0}+e^{2}A_{1}\approx
0.
\end{equation}
and
\begin{eqnarray}
\tilde{H}_{c}&=&\int dx [\frac{1}{2}\pi_{1}^{2}+\pi_{1}A_{0}^{\prime}+ \phi^{\prime2}
-2e\phi^{\prime}(A_{0}-A_{1})+\frac{1}{2}e^{2}(A_{0}-A_{1})^{2}\nonumber\\
&-&\frac{1}{2}ae^{2}(A_{0}^{2}-A_{1}^{2})],\label{ALHO}
\end{eqnarray}
respectively.  Note that with the gauge fixing conditions
(\ref{ALGF1}) and (\ref{ALGF2}) push back  the constraint $T_3$
into $\tilde{T}_{3}$ which was the constraint of the usual phase
space and as a result  $H_{c}$ lands onto $\tilde{H}_{c}$ which
was the hamiltonian of the usual phase space. It has therefore
become evident that physical contents remains the same in the
gauge symmetric version of the theory in the extended phase space.
The extra fields therefore renders there incredible service
towards bring back of the symmetry without disturbing the physical
sector.

For completeness of the analysis we compute the Dirac brackets of
the physical fields using the definition (\ref{DDB}). The matrix
$C_{ij}$ in this situation is
\begin{eqnarray}
C_{ij}=\left(\begin{array}{ccccc}
0 & 0 & 0 &0 & -e^{2}(a-1) \\
0 & -2\partial  & 0 & 0  & 0 \\
       0 & 0  & 0&-e^{2}\partial& 0 \\
       0 & 0 & -e^{2}\partial & 0& e^{2}\partial\\
        e^{2}(a-1)&0  &0 &e^{2}\partial & 0
\end{array}\right)\delta(x-y), \label{alter15}
\end{eqnarray}
and the inverse of it is the following
\begin{eqnarray}
C_{ij}^{-1}=\left(\begin{array}{ccccc}
0&0& \frac{1}{e^{2}(a-1)}\delta(x-y) & 0 & \frac{1}{e^{2}(a-1)}\delta(x-y) \\
   0& -\frac{\epsilon}{4}(x-y) & 0 & 0 &0  \\
       -\frac{1}{e^{2}(a-1)}\delta(x-y) & 0 &0 & -\frac{\epsilon}{2e^{2}}(x-y)& 0 \\
        0 & 0 & -\frac{\epsilon}{2e^{2}}(x-y) & 0& 0\\
             -\frac{1}{e^{2}(a-1)}\delta(x-y)&0 & 0 & 0 & 0\\
\end{array}\right).\label{ALTIM}
\end{eqnarray}
Using the definition (\ref{DDB}), it is straightforward to compute
the Dirac brackets between the field variables:
\begin{equation}
[A_{0}(x),A_{1}(y)]^{*}=\frac{1}{e^{2}(a-1)}\delta'(x-y),
\label{ADB1}
\end{equation}
\begin{equation}
  [\phi(x),\phi(y)]^{*}=-\frac{1}{4}\epsilon{(x-y)},\label{ADB2}
  \end{equation}
\begin{equation}
[A_{0}(x),\phi(y)]^{*}=\frac{1}{e(a-1)}\delta(x-y), \label{ADB3}
 \end{equation}
\begin{equation}
[A_{0}(x),\pi_{1}(y)]^{*}=-\frac{1}{(a-1)}\delta(x-y),
\label{ADB4}
\end{equation}
\begin{equation}
 [A_{0}(x),\pi_{\phi}(y)]^{*}=-\frac{1}{e(a-1)}\delta'(x-y), \label{ADB5}
 \end{equation}
 \begin{equation}
  [A_{1}(x),\pi_{1}(y)]^{*}=\delta{(x-y)}, \label{ADB6}
  \end{equation}
   \begin{equation}
  [\phi(x),\pi_{\phi}(y)]^{*}=\delta{(x-y)}, \label{ADB7}
  \end{equation}
Here also $(*)$ stands to symbolize the Dirac bracket. Note that
the Dirac brackets between the fields computed here are identical
with the set of Dirac brackets computed in Sec. II. It is indeed
the expected result.
\section{conclusion}
We have started our investigation with the gauged version of the
Siegel type chiral boson. From this action we have landed onto the
gauged version of FJ type chiral boson. Harada in \cite{KH} showed
that this action can be derived from JR version of Chiral
Schwinger model imposing a chiral constraint into the phase space
of the theory. Our investigation however reveals that the gauged
version of FJ type chiral boson is contained within the Seigel
action in an interesting way.  In fact, it is a successful
endeavor of obtaining the gauged version of chiral boson in a
different line of approach.

An extension towards the  BRST invariant reformulation of the
gauged version of the FJ type chiral boson has been made  using
BFV based improved FIK formalism. Though in \cite{SG}, an attempt
was made towards BRST quantization of the same model. However, in
that work the part of the effective action which was demanded as
the Wess-Zumino term did not agree with the Ref. \cite{KHWES}. In
spite of that, with that Wess-Zumino term the author established
the on-shell BRST invariance.

The way we have made the BRST invariant reformulation leads to the
appropriate Wess-Zumino term and this  does agree with the Ref.
\cite{KHWES}. It is interesting that the appropriate Wess-Zumino
term has automatically appeared during the process of BRST
quantization and with Wess-Zumino term, we observe the off shell
BRST invariance.

 An alternative
quantization has found possible due the appearance of the
appropriate Wess-Zumino term. From alternative quantization we
have seen that the theoretical spectrum agrees with the spectrum
obtained in the quantization of the gauge non-invariant version of
this model. It is indeed the expected result.

An equivalence between the gauge invariant version of the gauge
model of FJ type chiral boson of the extended phase space has been
established with the gauge non-invariant version of the usual
phase space following the same line of approach as it was
available from the work \cite{FALCK}. It is worth-mentioning that
the gauge fixing plays an important role to establish this
equivalence.
\section{Acknowledgements}
AR likes to thankfully acknowledge the provision of getting
computer facilities of Saha Institute of Nuclear Physics. He also
likes to thank Prof. Paul Bracken for providing reprints of few
works.


\begin{thebibliography}{the}
\bibitem{ST1} N. Marcur, J. Schwasz: Phys. Lett. {\bf B54} 111 (1982)
\bibitem{ST2} D. J Gross, J. A. Hervey, E. Martinec, R. Rohm: Phys. Rev. Lett.{\bf 54} 502 (1985)
\bibitem{ST3} C. Imbimbo, A. Schwimmer: Phys. Lett. {\bf B193} 455 (1987)
\bibitem{ST4} J. M. F. Labastida, M. Pernici: Nucl. Phys. {\bf B297} 557 (1988)
\bibitem{HAL1} S. G. Wen: Phys. Rev. Lett. {\bf 64} 2206 (1990)
\bibitem{HAL2} S. G. Wen: Phys. Rev. {\bf B41}, 12838 (1990)
\bibitem{SIG} W. Siegel: Nucl. Phys. {\bf B238} 307 (1984)
\bibitem{PPS} P. P. Sirvastava: Phys. Rev. Lett. {\bf 63}  2791 (1989)
\bibitem{FJ}  R. Floreanini,  R. Jackiw: Phys. Rev. Lett. {\bf 59}  1873  (1987)
\bibitem{JSON} J. Sonnenschem: Nucl. Phys. {\bf B309} 752 (1988)
\bibitem{RPM}D. Shukla, T. Bhanja, R. P. Malik: Eur. Phys. Jour. {\bf C74} 3025
(2014)
\bibitem{BPM}S. Upadhyay, B. P. Mandal: Eur. Phys. Jour. {\bf C71} 1759 (2011)
\bibitem{BEL} S. Bellucci, M. F. L. Golterman, D. N. M. Petcher: Nucl. Phys. {\bf B326} 307 (1989)
\bibitem{KH} K. Harada: Phys. Rev. Lett. {\bf 64} 139 (1990)
\bibitem{BAR} R. Amorim, J. Barcelos-Neto: Phys.Rev. D53 (1996) 7129-7137
\bibitem{BAL} D. Baleanu, Y. Guler: Czech.J.Phys. 52 (2002) 1171-1176
\bibitem{ARC} A. Rahaman: Int.J.Mod.Phys. {\bf A21} 1251 (2006)
\bibitem{ABREU} M.C. Abreu , A.  S. Dutra, C. Wotzasek: Phys.Rev. D67 (2003) 047701
\bibitem{EUNE} M. Eune, W. Kim E. J. Son: Eur.Phys.J. C71 (2011) 1840
\bibitem{SG} S. Ghosh: Phys. Rev. {\bf D49}, 2990 1994
\bibitem{JR} R. Jackiw, R. Rajaraman: Phys. Rev. Lett. {\bf 54}, 1219 (1985)
\bibitem{PM} P. Mitra: Phys. Lett. B284, 23 (1992)
\bibitem{PMS} S. Ghosh, P. Mitra: Phys. Rev. {\bf D44}, 1332 (1990)
\bibitem{ARPL} A. Rahaman: Phys. lett. {\bf B697} 260 (2011)
\bibitem{SM2} S. Mukhopadhyay, P. Mitra:  Ann.  Phys. (N. Y.){\bf 241} 68 (1995)
\bibitem{HAG}  C. R. Hagen, Ann. Phys. (N. Y.) {\bf 81} 67 (1973)
\bibitem{ARUP} A. Rahaman: Mod.Phys.Lett. {\bf A29}  1450072 (2014)
\bibitem{SCH} J. Schwinger: Phys. Rev. {\bf 128}, 2425 (1962)
\bibitem{ARAN1} A. Rahaman: Ann. Phys. (N. Y.) {\bf 354} 511 (2015)
\bibitem{ARAN2} A. Rahaman: Ann. Phys. (N. Y.) {\bf 361} 33 (2015)
\bibitem{BRS1} C. Becchi, A. Rouet,  R. Stora: Phys. Lett. {\bf B52} , 344 (1974)
\bibitem{BRS2}C. Becchi, A. Rouet,  R. Stora: Commun. Math. Phys. {\bf 42}, 127 (1975)
\bibitem{BRS3}C. Becchi, A. Rouet,  R. Stora: Ann. Phys. {\bf 98}  287 (1976)
\bibitem{BAT1} I. A. Batalin, V. Tyutin: Int.J. Mod. Phys. {\bf A6}, 3255 1991
\bibitem{BAT2} I. A. Batalin, E. S. Fradkin: Nucl. Phys. {\bf B279}, 514 1987
\bibitem{BA3} E. S. Fradkin,  G. A. Vilkovisky : Phys. Lett. {\bf B55}, 224 (1975)
\bibitem{BA4}I.A. Batalin,  V. Tyutin: Int. J. Mod. Phys. {\bf A6}, 3255 (1991)
\bibitem{FIK} T. Fujiwara, I. Igarashi,  J. Kubo: Nucl. Phys. {\bf B314}, 695 (1990)
\bibitem{WSJ} J. Wess, B. Zumino: Phys. lett. {\bf B37} 95 (1971)
\bibitem{KHWES} K. Harada I. Tsutsui: Phys. Lett. {\bf B183} 311 (1987)
 \bibitem{KIM1} Y. W. Kim,  S. K. Kim,  W. T. Kim,  Y.J. Park,  K. Y. Kim,  Y. Kim: Phys. Rev.
  {\bf D46}, 4574 1992
 \bibitem{KIM2} S.J. Yoon,  Y.W. Kim, Y.J. Park: J. Phys. {\bf G25}, 1783 1989
\bibitem{KIM3} Y. W. Kim, Y. J. Park S. J. Yoon: J.Mod.Phys. {\bf  A12} 4217 (1997)
\bibitem{MIAO} J. G. Zhou, Y. G. Miao, Y. Y Liu: J. Phys. {\bf G20}, 35
1994
\bibitem{PBRA} P. Bracken: Int. Jour. Theor. Phys. {\bf 47}, 3321 (2008)
\bibitem{AR1} A. Rahaman, S. Yasmin, S. Aziz: Int. Jour. Theor. Phys. {\bf 49}, 2607 (2010)
\bibitem{AR2} A. Rahaman: Int. Jour. Mod. Phys. {\bf A12}, 5625 (1997),
\bibitem{AR3} A. Rahaman:hep-th/9511097
\bibitem{AR4} S. Yasmin, A. Rahaman: Int.J.Mod.Phys.{\bf A31}  1650171 (2016)
\bibitem{DIR} P. A. M. Dirac: Lectures on Quantum Mechanics. Yeshiva University Press, New York 1964
\bibitem{SM1} S. Mukhopadhyay, P. Mitra:  Zeit. f.  Phys. {\bf C97} 552 (1995)
\bibitem{FALCK} N. C. Falck. G. Kramer:  Ann. Phys. (N. Y.) {\bf 176}
330 (1987)
\end{thebibliography}
\end{document}